\documentclass[preprint,12pt]{elsarticle}

\usepackage{amssymb}
\usepackage{amsmath}

\usepackage{xspace}
\usepackage{xcolor}
\usepackage{color}
\usepackage[normalem]{ulem}

\usepackage{lineno}

\newcommand{\dvlthgem}{\ensuremath{\Delta V_{\textrm{L-THGEM}}}\xspace}
\newcommand{\dvgthgem}{\ensuremath{\Delta V_{\textrm{G-THGEM}}}\xspace}

\newcommand{\eextraction}{\ensuremath{E_{\textrm{extraction}}}\xspace}
\newcommand{\edrift}{\ensuremath{E_{\textrm{drift}}}\xspace}
\newcommand{\qeeff}{\ensuremath{{\textrm{QE}}_{\textrm{eff}}}\xspace}
\newcommand{\frefl}{\ensuremath{f_{\textrm{refl}}}\xspace}
\newcommand{\freco}{\ensuremath{f_{\textrm{R}}}\xspace}
\newcommand{\epsone}{\ensuremath{\varepsilon_{\textrm{S1'}}}\xspace}
\newcommand{\epstwo}{\ensuremath{\varepsilon_{\textrm{S2}}}\xspace}
\newcommand{\epsext}{\ensuremath{\varepsilon_{\textrm{L-G}}}\xspace}
\newcommand{\epsgas}{\ensuremath{\varepsilon_{\textrm{G-THGEM}}}\xspace}
\newcommand{\Yone}{\ensuremath{Y_{\textrm{S1'}}}\xspace}
\newcommand{\Ytwo}{\ensuremath{Y_{\textrm{S2}}}\xspace}

\newcommand{\Wsc}{\ensuremath{W_{sc}}\xspace}
\newcommand{\Wi}{\ensuremath{W_i}\xspace}
\newcommand{\ntwo}{N\ensuremath{_2}\xspace}
\newcommand{\lntwo}{L\ntwo}
\newcommand{\am}[0]{$^{241}$Am\xspace}

\newcommand{\registered}{\ensuremath{^{\textrm{\textregistered}}}\xspace}

\newcommand{\ignoreblock}[1]{}

\journal{Nuclear Physics A}

\begin{document}

\begin{frontmatter}

\title{First studies on cascaded dual-phase liquid hole-multipliers in xenon}

\author[a,b]{G. Mart\'inez-Lema}
\ead{gonzalom@post.bgu.ac.il}

\author[a,b]{A. Roy}

\author[b]{A. Breskin}

\author[a]{L. Arazi}
\ead{larazi@bgu.ac.il}

\affiliation[a]{
organization={Unit of Nuclear Engineering, Ben Gurion University of the Negev},
city={Beer-Sheva},
country={Israel}    
}

\affiliation[b]{
organization={Department of Particle Physics and Astrophysics, Weizmann Institute of Science},
city={Rehovot},
country={Israel}
}

\begin{abstract}
Challenges in scaling up noble-liquid time projection chambers prompted the exploration of new detection concepts. The liquid hole-multiplier (LHM) was introduced as a potential component, enabling the detection of ionization electrons and VUV photons. Prior studies focused on perforated electrodes coated with CsI immersed in the liquid and electroluminescence amplification produced on a bubble trapped underneath. However, the performance was hindered by electron transfer across the liquid-gas interface. Here, we explored a bubble-free variant, placing a CsI-coated Thick Gas Electron Multiplier electrode below the liquid-gas interface to improve the transfer efficiency across it. Results show $>$5-fold improvement in the S1'/S2 ratio (a proxy for the photon detection efficiency (PDE)) compared to the bubble-assisted LHM. Although the achieved PDE is still below expectation ($\sim$4\%), we propose potential improvements to enhance the performance of this detector.

\end{abstract}



\begin{keyword}


Charge transport, multiplication and electroluminescence in rare gases and liquids \sep
Cryogenic detectors \sep
Micropattern gaseous detectors \sep
Noble liquid detectors \sep
Time projection Chambers \sep
\end{keyword}

\end{frontmatter}


\section{Introduction}
\label{sec:intro}

Dual-phase time projection chambers (TPCs) are among the leading technologies in the field of rare-event searches.
The current generation of dark matter (DM) experiments deploy tonnes of liquid xenon (LXe) or liquid argon (LAr) in cylindrical TPCs instrumented with photosensors.
The working principle relies on the detection of the primary scintillation light (``S1'') generated when a particle interacts in the active volume, and a delayed electroluminescence (EL) signal (``S2''), proportional to the number of ionization electrons released at the interaction, produced when the electrons are extracted to the gas phase under an intense electric field.

The TPC technology has proven highly successful in direct DM searches.
Earlier and current experiments searching for Weakly Interacting Massive Particles did not lead to the discovery of DM candidates \cite{PhysRevLett.131.041003,2022arXiv220703764A,PhysRevLett.129.161803}.
Therefore larger and more sensitive detectors are being conceived with expected masses of $\sim$50~tonne in LXe \cite{Aalbers_2016,Aalbers_2023} and 50 (DarkSide-20k \cite{Aalseth_2018}) and 370~tonne (ARGO \cite{Agnes_2021}) in LAr.
However, the expansion of the dual-phase TPC technology faces three major challenges.
One of them is the need to maintain a stable liquid-gas interface in-between and parallel to the gate and anode meshes, which are typically placed 5 to 10~mm apart.
At the scale of the proposed detectors ($>$2~m in diameter for LXe \cite{Aalbers_2016,Aalbers_2023}, $>$3~m for LAr in DarkSide-20k \cite{Aalseth_2018,Agnes_2021}), electrostatic forces between the meshes cause sagging which can introduce nonuniformities in the detector response.
This feature has a direct impact on the energy resolution and background-rejection capabilities of the detector.
Another concern is the ability to unambiguously detect S1 signals.
The detection of few-keV nuclear recoils requires sensitivity at the level of single photoelectrons, making it difficult to differentiate them from the intrinsic dark noise of the photosensors.
The larger the detector, the more photosensors are needed, exacerbating the problem.
This prevents the use of silicon photomultipliers (SiPMs), which are preferred to photomultiplier tubes (PMTs) due to their lower radioactivity, but have a much higher dark count rate \cite{ANFIMOV2021165162}.
Another important issue is the delayed electron emission from liquid to gas, which hampers low-energy S2-only analyses \cite{LUX_single_e_emission2020, Kopec_2021, XENON1T_single_e_emission2022}.

Much of the instrumentation-related research in the field of DM searches has been focused on finding solutions to these problems.
Several concepts have been proposed, both with single-phase and dual-phase configurations \cite{Breskin_2022}.
Among the single-phase concepts, thin wires \cite{Aprile_2014, Lin_2021_wires, Juyal_2021_wires, Wei_2022, Qi:2023bof} and microstrip plates \cite{POLICARPO1995568, Martinez-Lema:2023zjk} have been studied, as a way of generating electroluminescence and charge multiplication in the liquid.
Dual-phase concepts, such as the Liquid Hole-Multiplier (LHM) \cite{Breskin2013, Arazi:2015uja,
Erdal:2017yiu, Erdal:2018bjg, Erdal:2019dkk, Erdal:2019wfb}, and the Floating Hole Multiplier (FHM) \cite{Chepel_2023} also offer potential solutions to some of the aforementioned drawbacks of current dual-phase detectors.

The bubble-assisted LHM \cite{Erdal:2019dkk}, consists of a Gaseous Electron Multiplier (GEM \cite{SAULI1997531}) or a Thick GEM (THGEM \cite{BRESSLER2023104029}) coated with cesium iodide (CsI) UV-photocathodes immersed in the noble liquid, with a stable vapor bubble underneath.
In this concept, the detector is sensitive to both ionization electrons and scintillation photons induced by radiation within the noble liquid.
Scintillation photons (S1) extract photoelectrons that drift into the electrode’s holes; they cross the liquid-vapor interface into the bubble, where they induce
an EL signal (termed S1'). Similarly, the ionization electrons deposited in the liquid drift into the electrode’s holes and induce a delayed EL signal (S2) across the bubble. EL photons are recorded with nearby photosensors, providing the event’s deposited energy and localization. 
However, this technique suffered from a low photon detection efficiency (PDE).
Under optimal conditions, the LHM detector yielded $\sim$400 photons/e-/4$\pi$ in LXe. However, the low PDE values reached suggest electron losses induced, on the one hand, by somewhat low photo-electron collection efficiency into the electrode's holes and possibly an inefficient transmission of electrons through the liquid-gas interface into the bubble \cite{Erdal:2017yiu,Tesi_2021}. The latter must have also affected the energy resolutions ($\sim$6\% for S1 and S2 with 5.5~MeV alpha particles).

Here we explore a new variant of the LHM concept using a flat liquid-gas interface located above the immersed LHM, as proposed in \cite{Martinez-Lema:2023qox}: the cascaded dual-phase LHM.
We investigated two implementations of this idea: (1) a single-THGEM configuration, 
with a CsI-coated THGEM immersed in LXe and EL occurring in a parallel gap in the vapor phase,
and (2) a double-THGEM, 
with the CsI-coated electrode immersed in the liquid and EL occurring within the holes of a second THGEM placed in the vapor phase.
Note that a similar 2-THGEM configuration was previously explored in LAr, in a different way, for detecting S2 electrons.
The double-THGEM configuration offers potential solutions to the aforementioned drawbacks of current dual-phase detectors.
The CsI-coated THGEM immersed in the liquid followed by EL within the holes of a second perforated electrode have the potential to enhance the detection of S1 signals, as single photoelectrons released from the photocathode would produce a flash of light.
These flashes will be unambiguously detected, as they produce signals far above the photosensors' dark noise.
The introduction of a second THGEM in the gas phase decouples the processes of electron extraction to the gas phase and EL amplification.
This has two main benefits: (1) there is virtually no sagging, as THGEMs are much stiffer than meshes, allowing the employment of higher extraction field, which decreases the contribution of delayed electron emission; (2) it confines the EL generation to a small region, away from the liquid-gas interface, which reduces the sensitivity to imperfections at the liquid-gas interface.

In what follows, we describe the operation principles and performance of the single- and double-THGEM configurations with an emphasis of the gain in PDE compared to that of the bubble-assisted LHM.

\section{Experimental setup}
\label{sec:setup}

The measurements reported in this work were performed in the MINiX cryostat, described in detail in \cite{Erdal:2015kxa}.
This dedicated LXe cryostat consists of a cylindrical chamber 100~mm in diameter and 100~mm in height, accommodating around 0.5~L of LXe.
Thermal insulation is provided by an outer vacuum-sealed volume, while the inner space houses the detector assembly.
Suspended from the top flange, the assembly includes high-voltage feedthroughs to bias the electrodes.
A fused-silica viewport, positioned at a $60^{\circ}$ angle from the vertical axis, enables visual inspection of the assembly.
A section of the cryostat wall, temperature-modulated by \lntwo cooling, allows for a controlled gas liquefaction.
Continuous xenon purification is performed using a recirculation system.
LXe is extracted from the bottom of the cryostat directly through a heat exchanger to a SAES hot getter (model PS3-MT3-R-2).
The purified xenon is reintroduced into the upper part of the cryostat via the heat exchanger.

To protect the CsI photocathode from degradation, the detectors were assembled in an \ntwo environment.
The CsI photocathode was vacuum-deposited onto the THGEM electrode in a dedicated evaporator.

Before filling with gaseous Xe, the chamber underwent pumping down to approximately $10^{-6}$~mbar. The reported experimental results were obtained following several days of xenon recirculation at a rate of $\sim$1.2 slpm while maintaining a constant LXe temperature of 175~K ($\pm 0.1$~K) throughout the measurements.

\begin{figure*}
    \centering
    \includegraphics[height=52mm]{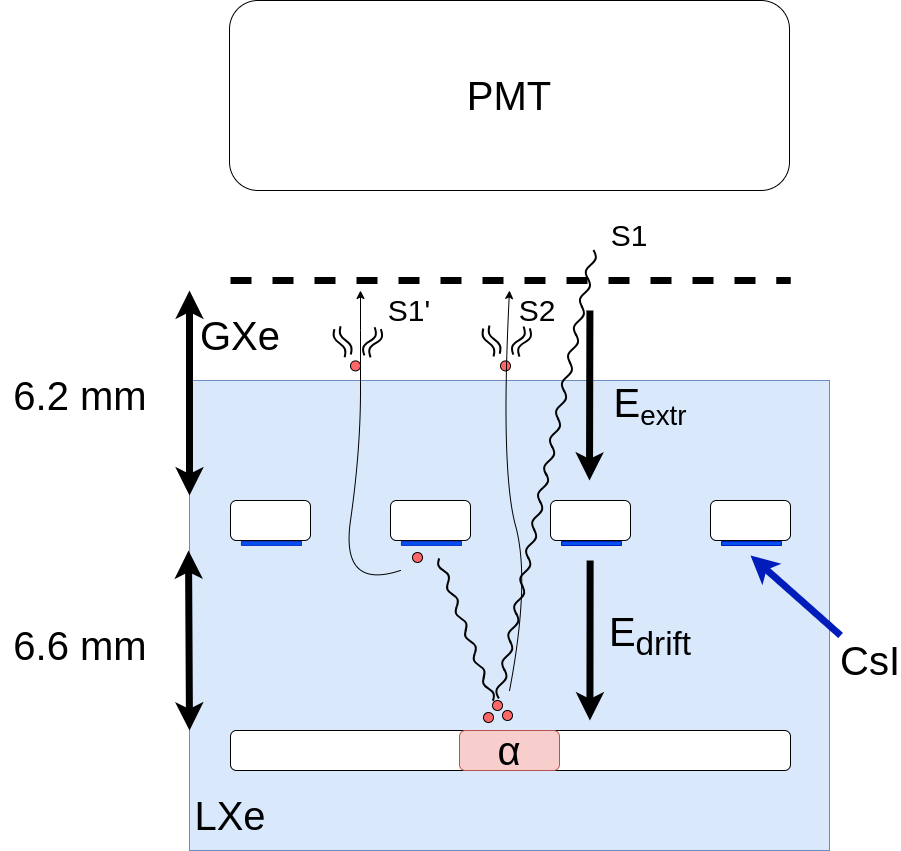}
    \includegraphics[height=52mm]{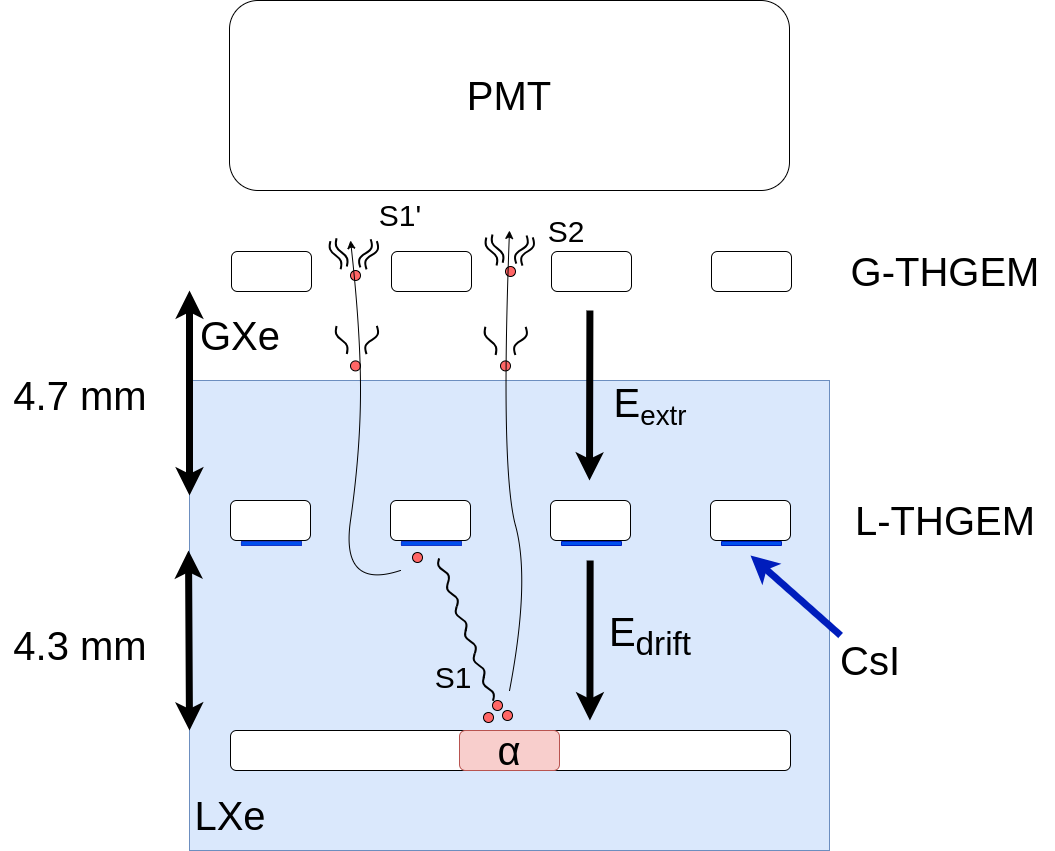}
    \caption{Schematic drawing of the setups used in our experiments. Left: single-THGEM configuration. Right: dual-THGEM configuration.
    Both ionization electrons and scintillation-induced photoelectrons from CsI are collected into the L-THGEM holes; they are extracted from the liquid into the vapor phase. EL occurs in a parallel gap (left) or within the G-THGEM holes (right).}
    \label{fig:setups}
\end{figure*}

We used two different detector configurations, depicted in Figure \ref{fig:setups}.
The first setup (left panel), labeled ``single-THGEM'' configuration, consisted of a THGEM coated with CsI on its bottom face, immersed in LXe (labeled ``L-THGEM'') and a mesh placed 6.2~mm above the liquid-gas interface.
The CsI coating was a circular patch 450-nm thick, deposited in the central L-THGEM area with a diameter of 20~mm.
Below the L-THGEM, an \am alpha source was placed in a stainless steel frame that acts as a cathode.
The mesh consisted of 50~$\mu$m stainless-steel wires woven to form a square pattern with a pitch of 500~$\mu$m, providing a transparency of 81\%.
For the second setup (right panel), the mesh was replaced by a second THGEM (``G-THGEM''), without CsI coating located in the vapor phase 4.7~mm above the L-THGEM.
A few mm above the mesh or the G-THGEM, we placed a square 1-inch VUV-sensitive PMT (Hamamatsu R8520-406) to record the light produced in the events.
The distance between the liquid-gas interface and the L-THGEM could not be determined directly.

Both setups used 33-mm-diameter, gold-plated THGEM electrodes made of 0.4~mm thick FR4, with 0.3~mm diameter holes in a hexagonal pattern and a pitch of 0.7~mm for both the L- and the G-THGEMs.
The holes in the insulator were surrounded by 50~$\mu$m wide rims.
All the elements in the setup were mounted on PEEK rods to ensure insulation from the cryostat body.
The gaps between the different components were created using either PEEK or FR4 spacers.
CAEN N471A power supplies were used to bias the electrodes.
The PMT waveforms were digitized and recorded with a Tektronix 5204B oscilloscope, with waveform analysis performed offline.
The currents measurements reported in Section \ref{sec:ete} were recorded with a Keithley 610C solid-state analog electrometer.

\section{Waveform analysis}
\label{sec:analysis}

The measurements presented 
were carried out under several voltage configurations.
For each measurement, the digitized waveforms were integrated in fixed-width time windows to obtain the integrated charge.
The position of these windows was adjusted according to the voltage configuration in each measurement.
An in-situ PMT calibration -- following the procedure detailed in \cite{DOSSI2000623} -- allowed us to translate the integrated charge into a number of photoelectrons.

We use the following notation:
\edrift for the electric field strength between the cathode and L-THGEM (taking into account the dipole field produced near the L-THGEM holes);
\dvlthgem for the voltage difference between the top and bottom faces of the L-THGEM;
\eextraction for the electric field strength in the liquid between the L-THGEM and the mesh or between the L- and the G-THGEM;
and \dvgthgem for the voltage difference between the top and bottom faces of the G-THGEM.

\begin{figure*}
    \centering
    \includegraphics[width=0.9\textwidth]{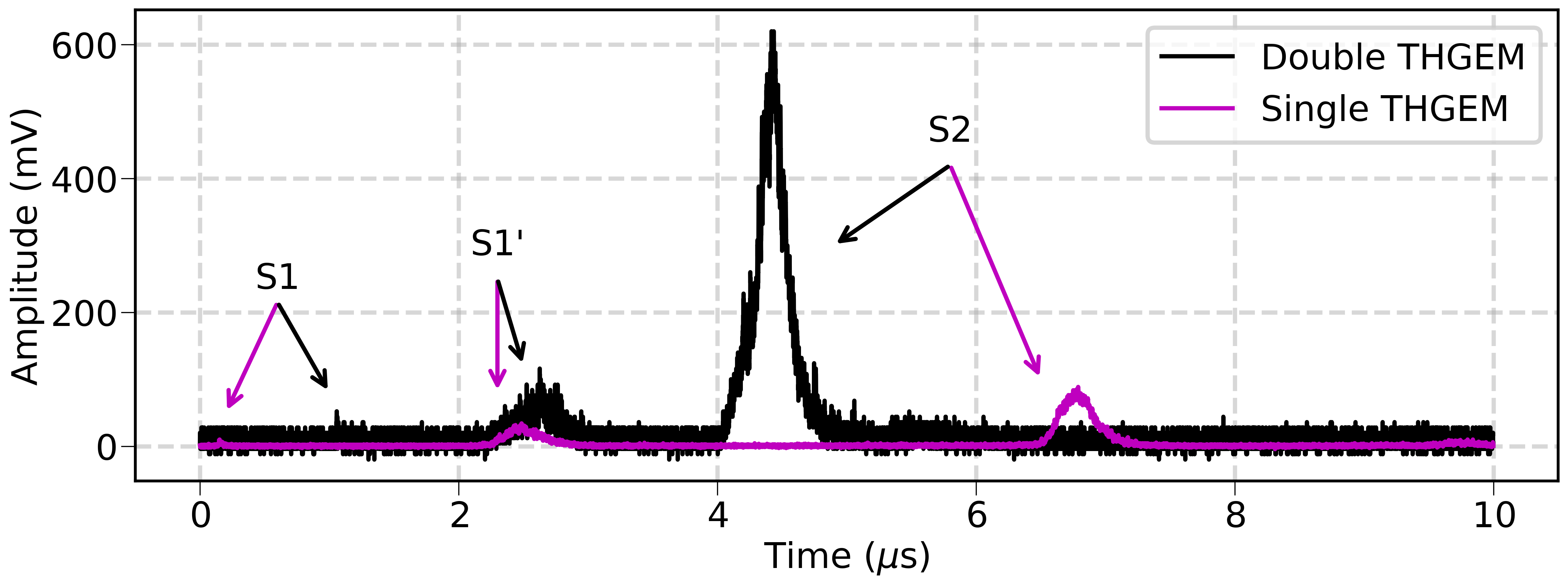}
    \caption{Typical PMT waveforms for an $\alpha$-particle interaction in the single-THGEM (purple) and double-THGEM (black) configurations shown in Figure \ref{fig:setups}.
    The S1, S1' and S2 pulses are labeled (see text for details).
    The different delays of S1' and S2 signals correspond to the different drift lengths in each setup.
    The waveforms were taken at \edrift = 660 V/cm, \dvlthgem = 2 kV, \eextraction = 6.8 kV/cm and \dvgthgem = 1.4 kV.}
    \label{fig:wf_comparison}
\end{figure*}

Figure \ref{fig:wf_comparison} shows a typical PMT waveform for both setups under similar voltage configurations.
A small fraction of the primary scintillation light produced in the alpha-particle interaction escapes through the holes of the THGEM(s) providing the S1 signal.
Another fraction of the primary scintillation light impinges on the CsI photocathode, releasing photoelectrons, which are focused into the L-THGEM holes.
An intense field above the L-THGEM extracts the photoelectrons to the gas phase, where they induce electroluminescence (EL) leading to the S1' signal.
The ionization electrons released in the interaction drift towards the L-THGEM and are focused into the holes.
They are extracted to the vapor phase to produce a second EL signal labeled S2.
The difference in time intervals between S1, S1' and S2 signals result from the different drift and extraction gaps in the two setups.

\section{Photon detection efficiency}
\label{sec:pde}

\subsection{Methodology}
\label{sec:methodology}

A key characteristic of LHM detectors is that the prompt scintillation light is detected as an amplified EL signal S1'; namely, a photoelectron induced by a single photon impinging on the CsI photocathode in the liquid phase will result in the detection of several EL-induced photons in the vapor phase \cite{Erdal:2015kxa}.
The photon detection efficiency (PDE) in our LHM geometries reflects the probability that a photon impinging the CsI photocathode is successfully detected by a photosensor recording EL photons.
This can be expressed as:


\begin{equation}
    PDE = \qeeff \cdot \epsone \cdot \epsext \cdot \epsgas,
    \label{eq:pde}
\end{equation}

\noindent
where \qeeff is the effective quantum efficiency of the photocathode, i.e. the average probability that a photoelectron is successfully extracted from CsI into the liquid; \epsone is the electron transfer efficiency across the L-THGEM holes for photoelectrons; \epsext is the efficiency of electron extraction to the vapor phase; and \epsgas is the electron focusing efficiency into the G-THGEM holes in the double-THGEM configuration (for the single-THGEM one, $\epsgas=1$).
The PDE can also be expressed in terms of the S1'/S2 ratio.
The integrated charge of S1' and S2 signals can be expressed as: 

\begin{equation}
    S1' = \frac{E}{\Wsc} \cdot (1 + \frefl) \cdot \frac{\Omega}{4\pi} \cdot PDE \cdot \Yone \\
    \label{eq:s1p}
\end{equation}
\begin{equation}
    S2  = \frac{E}{\Wi} \cdot \freco \cdot \epstwo \cdot \epsext \cdot \epsgas \cdot \Ytwo
    \label{eq:s2}
\end{equation}

\noindent
where $\Wsc = (17.1 \pm 1.4)$~eV \cite{CHEPEL2005160} and $\Wi = (15.6 \pm 0.3)$~eV \cite{PhysRevA.12.1771} are the average energy values expended in producing a primary scintillation photon and ionization electron in LXe, respectively; $\frefl$ is the reflectivity of the gold surface coating the alpha-source; \freco is the fraction of electrons escaping recombination \cite{APRILE1991119}; $\Omega$ is the solid angle subtended by the photocathode, accounting for the fraction of active CsI area (i.e., excluding the L-THGEM holes); \epstwo is the electron transfer efficiency across the L-THGEM for ionization electrons; and \Yone and \Ytwo are the average number of photons detected per electron in the gas phase for S1' or S2 signals, respectively.
Combining Eqs \ref{eq:pde}, \ref{eq:s1p} and \ref{eq:s2} we obtain:

\begin{equation}
    PDE = \frac{\Wsc}{\Wi}  \cdot  \frac{\freco \cdot \epstwo \cdot \epsext \cdot \epsgas}{(1 + \frefl) \cdot \frac{\Omega}{4\pi}} \cdot \frac{\Ytwo}{\Yone} \cdot \frac{S1'}{S2}
    \label{eq:pdes1ps2}
\end{equation}

\subsection{Measurement of the S1'/S2 ratio}
\label{sec:s1ps2ratio}

\begin{figure}
    \centering
    \includegraphics[width=0.49\columnwidth]{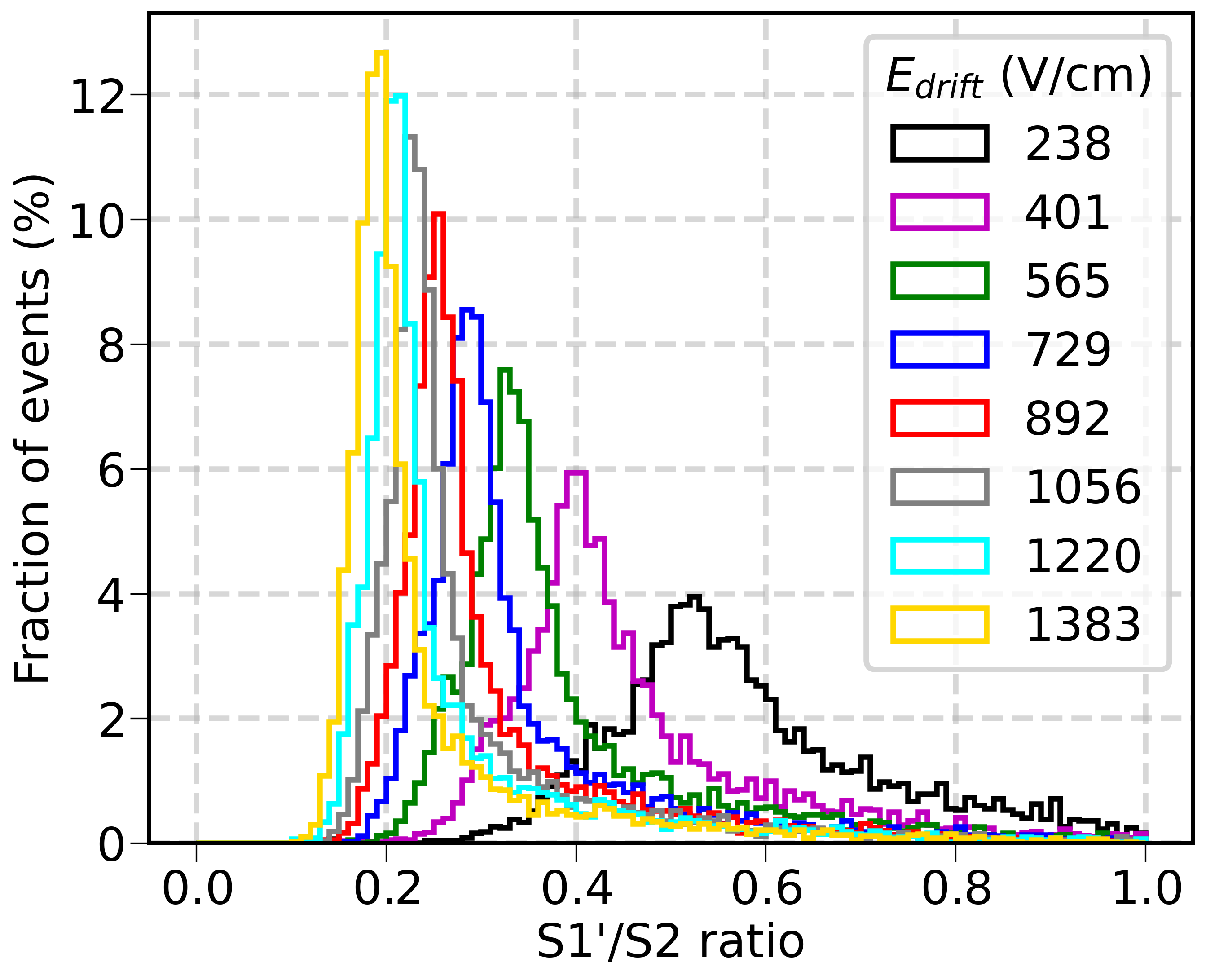}
    \includegraphics[width=0.49\columnwidth]{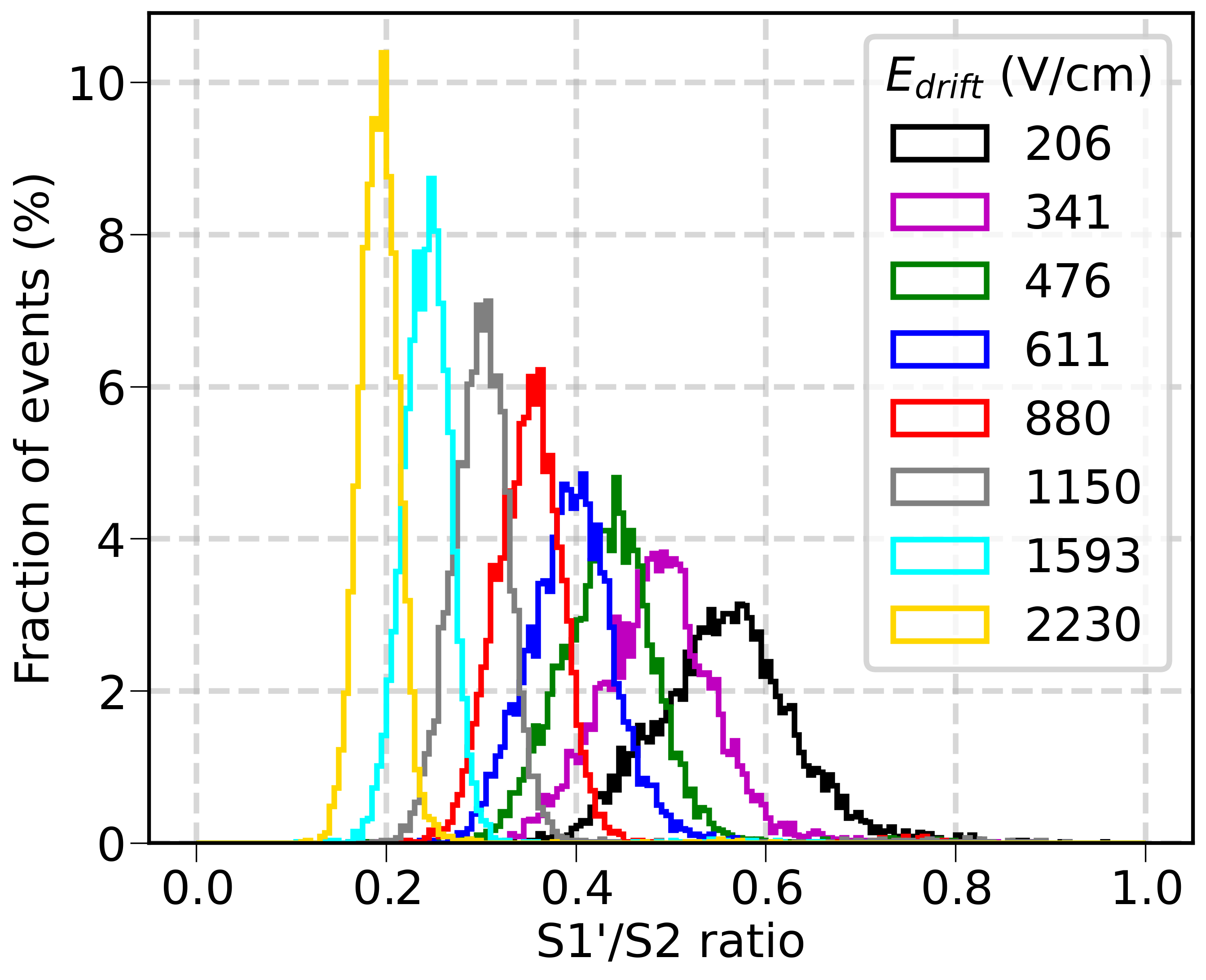}
    \caption{Distribution of the S1'/S2 ratio for various drift fields in the single-THGEM (left) and double-THGEM (right) configurations.}
    \label{fig:s1p_s2_ratio}
\end{figure}

The S1'/S2 ratio was measured for both configurations under similar conditions.
Figure \ref{fig:s1p_s2_ratio} shows the distributions of the ratio as a function of the drift field for the single-THGEM (left) and double-THGEM (right) configurations.
In both cases, \dvlthgem~=~2~kV and \eextraction~=~6.4~kV/cm; for the double-THGEM configuration \dvgthgem~=~1~kV.
We observe that the ratio increases as the drift field decreases in both configurations.
In addition, for a similar drift field, the S1'/S2 ratio is higher in the double-THGEM than in the single-THGEM configuration.

\begin{figure}
    \centering
    \includegraphics[width=0.7\columnwidth]{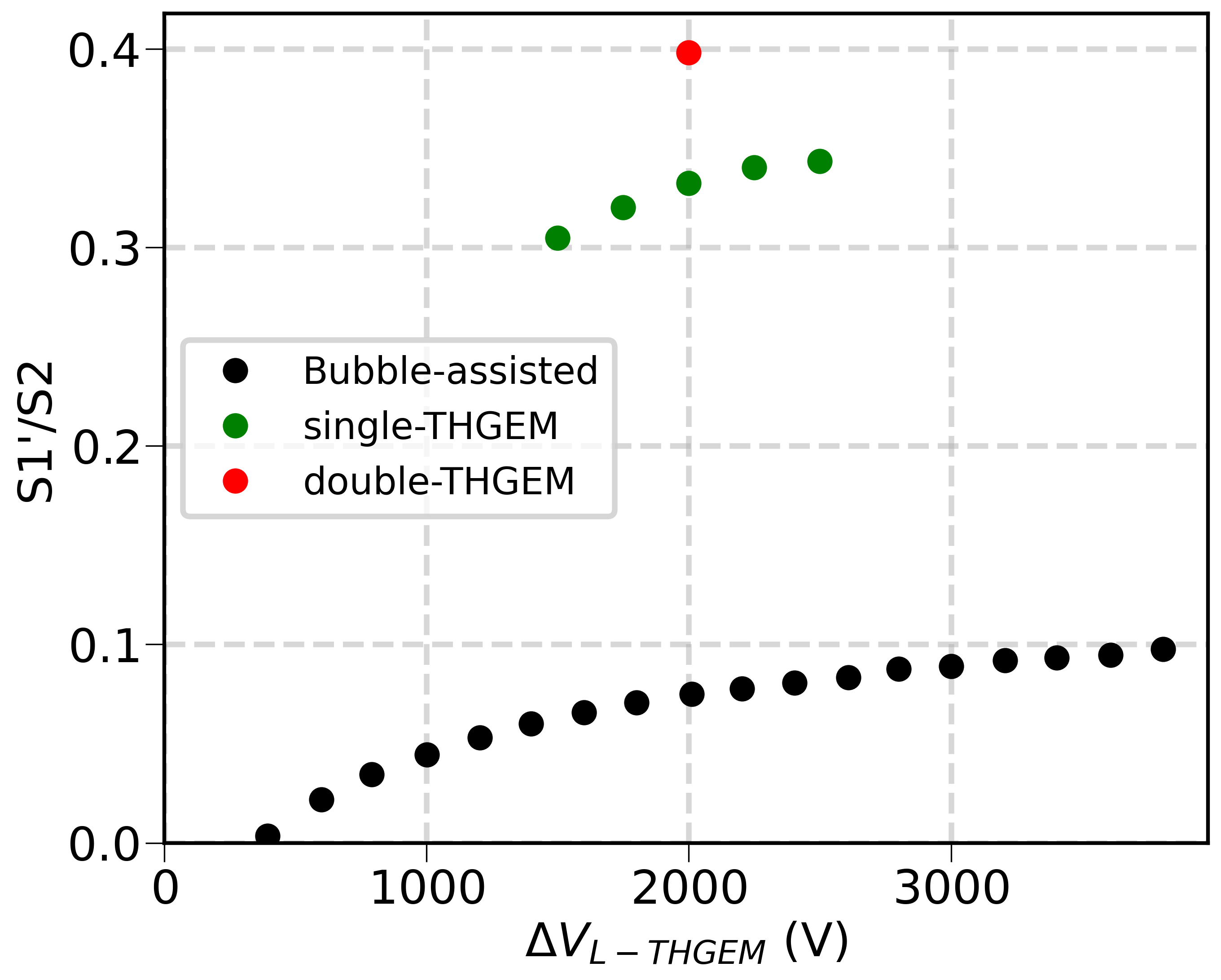}
    \caption{Comparison between the S1'/S2 ratio measured as a function of \dvlthgem in the bubble-assisted LHM (black points, reproduced from \cite{Erdal:2017yiu}), and in the single-THGEM (green points) and double-THGEM (red point) configurations.
    All measurements have been obtained under similar voltage configurations (see text for details).
    }
    \label{fig:s1p_s2_ratio_comparison}
\end{figure}

The S1'/S2 ratio was also measured for the single-THGEM configuration at \edrift = 729~V/cm as a function of \dvlthgem.
Figure \ref{fig:s1p_s2_ratio_comparison} shows a comparison between the S1'/S2 ratio measured in a bubble-assisted THGEM LHM \cite{Erdal:2017yiu} and in the two configurations of the cascaded THGEM LHMs.
The data reproduced from \cite{Erdal:2017yiu} corresponds to a drift field of 500~V/cm and the cascaded-LHM data to the closest values available: 565~V/cm and 540~V/cm for the single- and double-THGEM configurations, respectively.
We assume that the dependence on \edrift is the same for all values of \dvlthgem.
The S1'/S2 ratio measurements at various THGEM voltages have been scaled according to the ratio

$$\frac{R|_{\edrift = 565~V/cm}}{R|_{\edrift = 729~V/cm}}$$
where $R$ is the most probable value of the S1'/S2 distribution at \dvlthgem~=~2~kV.

The S1'/S2 values obtained in both configurations of the cascaded LHM are significantly higher than that obtained in the bubble-assisted THGEM LHM using an electrode of the same parameters and under similar conditions.
We obtain improvements of factors 4.4 and 5.3 at \dvlthgem~=~2~kV for the single- and double-THGEM configurations, respectively.
The S1'/S2 values are also higher than the best one obtained in the bubble-assisted LHM, with a single-conical GEM immersed in the liquid.

The main difference between the configurations studied here and the bubble-assisted LHM is the nature (location, size, and shape) of the liquid-gas interface.
While in previous studies the interface was associated with the bubble and confined to the LHM holes, here it is unconfined and located above the L-THGEM electrode, without contact with its surface.
The main hypothesis invoked in \cite{Erdal:2017yiu} to explain the low S1'/S2 ratio was electron loss on the bubble interface.
Indeed, a higher loss of S1' photoelectrons than of S2 ionization electrons would translate into a low S1'/S2 ratio.
Using the field lines as a proxy for the electron trajectories, photoelectrons from CsI reach the bubble liquid-gas interface within the LHM hole closer to the wall compared to S2 ionization electrons which are focused to the hole center, resulting in higher rate of loss to the wall. If the bubble interface is curved upward, i.e. reaches a maximal height at the hole center, the electric field vector has a larger tangential component relative to the interface closer to the wall; this may further increase the loss of S1' photoelectrons compared to S2 ionization electrons by pushing them towards the wall.
It is also known that, even though the bulk of electrons are transferred to the gas phase in a time scale of 1 ns or less, a fraction of the electrons are thermalized at the boundary \cite{BOLOZDYNYA1999314} and are emitted in time scales of tens of $\mu$s or ms \cite{Sorensen_2018}.
It is conceivable that the fraction of thermalized electrons depends on the curvature of the interface, which could potentially explain the difference in transfer efficiency to the bubble between photoelectrons and ionization electrons.

\subsection{Derivation of the PDE}
\label{sec:pde_derivation}

Based on the values of the refractive index of gold obtained by \cite{Werner2009}, we assume $f_{refl} = 0.28$ in Equation \ref{eq:pdes1ps2}.
The solid angle was calculated to be $\Omega = 2.11$~sr (16.8\%) and $\Omega = 2.66$~sr (21.2\%) in the single- and double-THGEM configurations, respectively.

The ratio \Ytwo/\Yone is not strictly one due to the different spatial distribution of S1' and S2 signals.
While the former can occur across the active area of the CsI photocathode ($\sim20$~mm diameter), S2 signals are distributed over the active area of the \am source ($\sim6$~mm diameter).
The light collection efficiency could in principle vary across these areas.
However, the region above the mesh/G-THGEM was surrounded by PTFE, which has good reflectivity.
To estimate the difference, we simulated the setup in Geant4 and compared the light detection efficiency for an on-axis source and for a source distributed homogeneously over the 20~mm diameter area covered by the photocathode.
Assuming a reflectivity of 72\% at 172 nm \cite{Silva:2010uhc}, we found that $\Ytwo/\Yone \approx 1.1$.

\epstwo was previously measured in \cite{Martinez-Lema:2023qox} and it is approximately 1 for the voltage configurations used in our experiments.
The experimental measurements of \epsext show some degree of variability across the literature.
Based on the results of \cite{Xu:2019dqb}, we assume \epsext = 0.9.
As discussed in Section~\ref{sec:alignment}, the value of \epsgas for the double-THGEM configuration depends strongly on the unknown alignment between the L- and G-THGEMs.
We will adopt a value of \epsgas = 0.4 based on the value reported in Section \ref{sec:ete} in absence of a better reference.
Naturally, for the single-THGEM configuration \epsgas~=~1.

The drift field has a strong impact on the effective quantum efficiency of extracting an electron from the photocathode, which affects directly the S1'/S2 ratio, but also changes significantly the fraction of ionization electrons escaping recombination.
We interpolate the data in \cite{APRILE1994328} to obtain a continuous function of \freco as a function of the drift field.

\begin{figure}
    \centering
    \includegraphics[width=0.7\columnwidth]{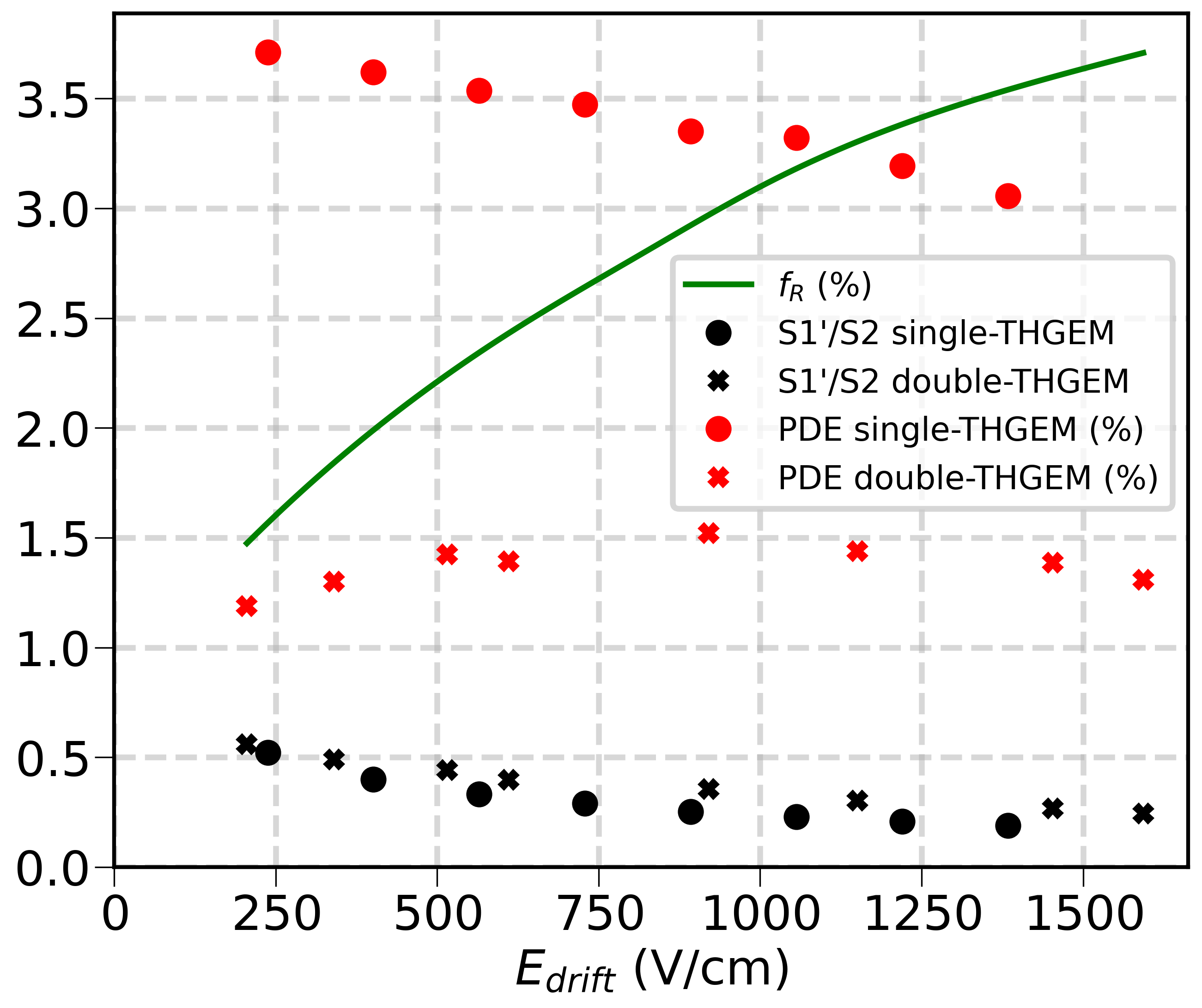}
    \caption{S1'/S2 ratio (black) and resulting PDE (red) as a function of \edrift for the single-THGEM (circles) and double-THGEM (crosses) configurations.
    The values of $f_R$ (green curve, reproduced from \cite{APRILE1991119}) are also displayed for reference.
    }
    \label{fig:pde}
\end{figure}

Figure \ref{fig:pde} shows the S1'/S2 ratio (black) and the resulting PDE (red) as a function of \edrift for the single-THGEM (circles) and double-THGEM (crosses) configurations.
The values of \freco (green curve) are also shown for reference.

The maximal PDE in each configuration is observed at different drift fields.
For the single-THGEM configuration, a value of 3.7\% is observed at \edrift~=~238~V/cm, while
a value of 1.5\% is obtained at \edrift~=~920~V/cm in the double-THGEM configuration.
The PDE in the single-THGEM configuration is notably higher than in the double-THGEM one.
This is attributed to the poor transfer efficiency across the G-THGEM due to a misalignment between the L- and G-THGEMs, as discussed in Section \ref{sec:alignment}.
This is reflected in the factor \epsgas in Eqs. \ref{eq:pde} and \ref{eq:pdes1ps2} and the ratio S1'/S2 in Eq. \ref{eq:pdes1ps2}.
A good alignment between both THGEMs can result in a nearly perfect transfer efficiency ($\epsgas \approx 1$).
Extrapolating the double-THGEM PDE to a setup with \epsgas~=~1 yields a PDE value in agreement with that of the single-THGEM configuration.

The PDE can also be obtained from Eq. \ref{eq:s1p}.
In the single-THGEM case we observe an average of $S1' = 1.4 \cdot 10^3$ PE at \edrift~=~238~V/cm.
This, combined with \Yone~=~0.66 PE/electron from Section~\ref{sec:light_yield}, yields a PDE value of 3.6\%.
For the double-THGEM, $S1' = 4.4 \cdot 10^3$ PE at \edrift~=~920~V/cm and \Yone~=~5.1 PE/electron, resulting in a PDE value of 1.2\%.
Both estimations are in agreement with the values obtained via the S1'/S2 ratio.

The PDE values obtained in the cascaded LHM are undoubtedly higher than that of the bubble-assisted THGEM LHM \cite{Erdal:2019dkk}, regardless of the configuration.
We attribute this to the higher S1'/S2 ratio attained in these setups.
Furthermore, for the single-THGEM configuration we obtain a PDE similar to that of the bubble-assisted LHM with a single-conical GEM \cite{Erdal:2019Thesis}.

\section{Comparison between single- and double-THGEM configurations}
\label{sec:comparison}

\subsection{Light yield}
\label{sec:light_yield}

\begin{figure}
    \centering
    \includegraphics[width=0.7\columnwidth]{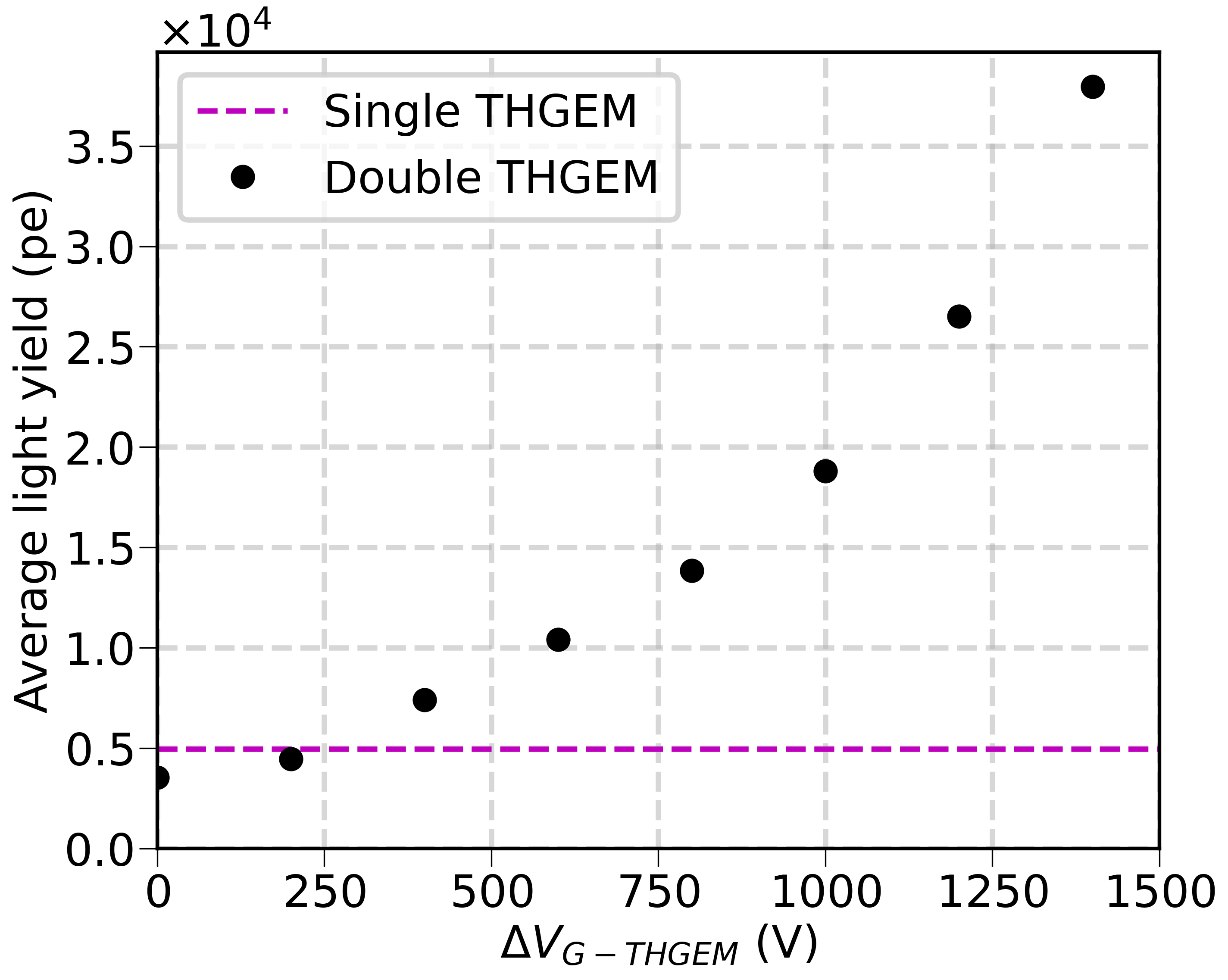}
    \caption{Measured light yield in the double-THGEM configuration (black points) as a function of \dvgthgem compared to the light yield obtained with the single-THGEM configuration (purple line) for \edrift~=~660~V/cm, \dvlthgem~=~2~kV and \eextraction~=~5.4~kV/cm.
    The light yield is increased by a factor 7.7 for \dvgthgem~=~1.4~kV.
    }
    \label{fig:light_yield}
\end{figure}

As shown in Figure \ref{fig:wf_comparison}, for the same values of \edrift, \dvlthgem and \eextraction, the double-THGEM configuration yields significantly larger S2 pulses.
This is confirmed in Figure \ref{fig:light_yield}, where we display the average of the S2 integral as a function of \dvgthgem in the double-THGEM setup (black points) and for the single-THGEM configuration (magneta line,  representing a constant value in the parallel gap, under the constant  extraction field).
Both measurements were performed under the same field configuration: \eextraction~=~5.4~kV/cm, \dvlthgem~=~2~kV and \edrift~=~660~V/cm.

A maximum light yield of $3.8 \cdot 10^4$ photoelectrons (PEs) was observed at \dvgthgem = 1.4~kV, a factor 7.7 greater than the one obtained in the single-THGEM configuration.
Assuming $7.5\cdot10^3$ electrons escaping recombination from an alpha-particle track ($N_e=E/W_i\cdot f_R$), this corresponds to approximately 5.1 detected photons per drifting electron.
For comparison XENON1T has 11.5~PE/e \cite{XENON:2017lvq}. Note that the G-THGEM voltage was limited to 1.4~kV due to discharges (likely on the feedthroughs, which were operated at a maximal potential of 1.6 kV to maintain the overall potential ladder across the entire setup).
    
\subsection{Energy resolution}
\label{sec:eres}

\begin{figure}
    \centering
    \includegraphics[width=0.7\columnwidth]{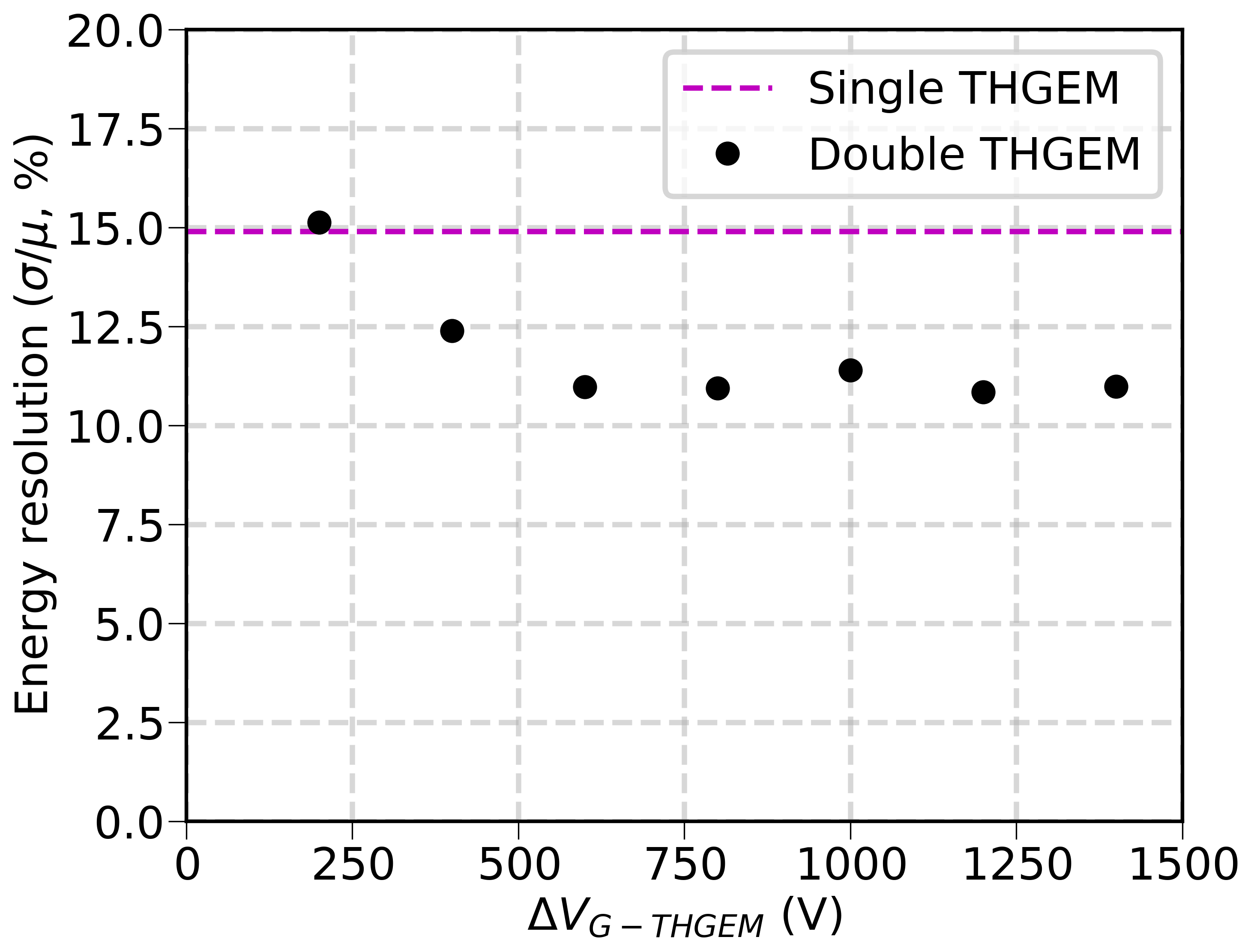}
    \caption{Measured energy resolution in the double-THGEM configuration (points) as a function of \dvgthgem for three different extraction fields.
    An energy resolution of $\sigma/E \sim$~11~\% is obtained for \eextraction~=~5.4~kV/cm and  \dvgthgem~=~1.4~kV.
    The best energy resolution achieved in the single-THGEM arrangement (dashed line) is also included for comparison.}
    \label{fig:eres_comparison}
\end{figure}

The energy resolution was determined by fitting the high-energy side of the S2 energy spectrum to a Gaussian distribution with mean $\mu$ and standard deviation $\sigma$.
Figure \ref{fig:eres_comparison} displays the energy resolution, defined as $\sigma/\mu$, measured in the double-THGEM setup as a function of \dvgthgem compared to the energy resolution measured in the single-THGEM setup under the same field configuration: \dvlthgem~=~2~kV, \eextraction = 5.4~kV/cm and \edrift~=~660~V/cm.
We observe that overall, the energy resolution of the double-THGEM configuration is superior to that of the single-THGEM.
Higher \dvgthgem values result in better energy resolution, in accordance to the increase in photoelectron statistics up to \dvgthgem = 600 V.
However, it saturates above this value.
The reader is referred to Section \ref{sec:discussion} for the discussion of this result.

\subsection{Pulse width}
\label{sec:pulse_width}

\begin{figure*}
    \centering
    \includegraphics[width=0.9\textwidth]{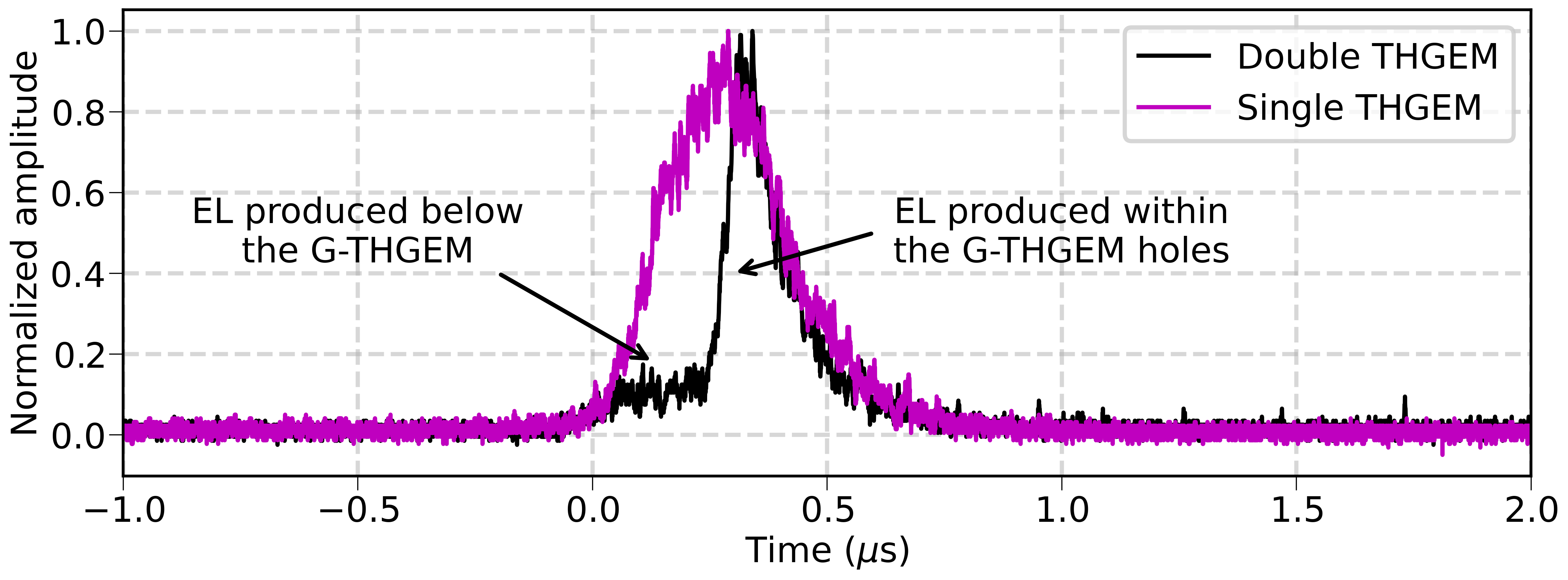}
    \caption{Amplitude-normalized S2 signals for the single-THGEM (left) and double-THGEM (right) configurations.
    The waveforms are synchronized such that both S2s start at the same time.
    The voltage configuration was \edrift = 660 V/cm, \dvlthgem = 2 kV, \eextraction = 6.8 kV/cm and \dvgthgem = 0.6 kV.
    }
    \label{fig:s2_comparison}
\end{figure*}

Figure \ref{fig:s2_comparison} shows a partial waveform recorded in each configuration for which the pulse amplitude was normalized to 1 and the S2 signals were synchronized.
We observe narrower S2 signals in the double-THGEM configuration.
Moreover, we observe that the double-THGEM S2 pulse is composed of two different contributions; a major fast one originating from the holes and a small slower one produced by EL in the uniform field underneath.
In the single-THGEM case, the EL light is produced roughly uniformly between the interface and the mesh.
In the double-THGEM configuration, most of the light is produced in the holes of the G-THGEM, as the electric field is much more intense.
However, a small fraction of light is still produced in the uniform-field regime, but is also partially blocked by the G-THGEM, diminishing its contribution even further.

\subsection{Correlation between S1' and S2}
\label{sec:correlation}

\begin{figure*}
    \centering
    \includegraphics[width=0.49\textwidth]{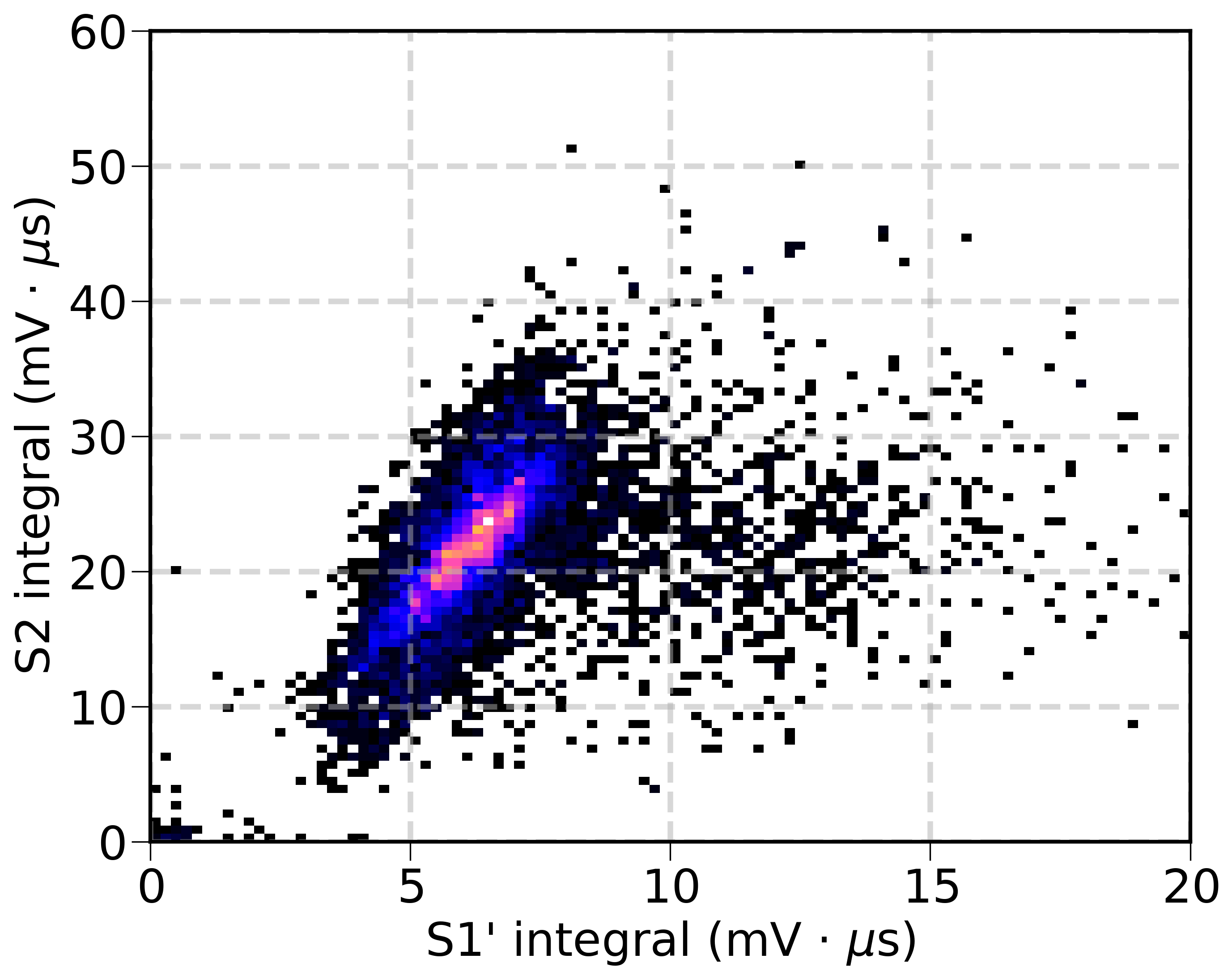}
    \includegraphics[width=0.49\textwidth]{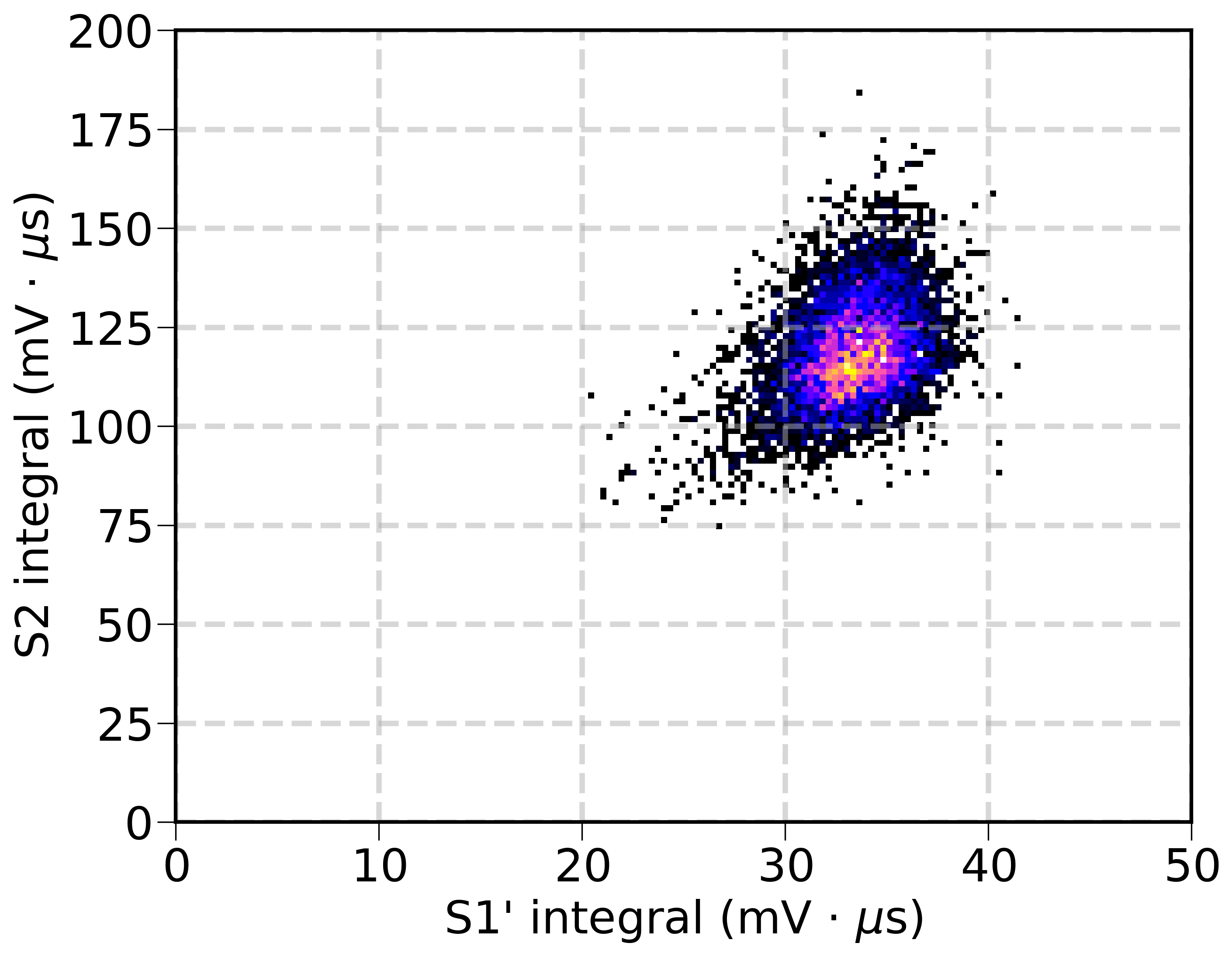}
    \caption{Integral of the S2 signal as a function of the integral of the S1' signal for the single-THGEM (left) and double-THGEM (right) configurations.
    The color scale denotes event density.
    The data was taken at \edrift~=~680~V/cm, \dvlthgem~=~2~kV, \eextraction~=~6.4~kV/cm, and (in the double-THGEM configuration) \dvgthgem~=~1~kV.}
    \label{fig:s1p_s2_correlation}
\end{figure*}

In Figure \ref{fig:s1p_s2_correlation} we show the distribution of the S2 vs S1' signals for the single-THGEM (left) and double-THGEM configurations (right).
We observe a significant correlation between them.
Besides the higher photon yields obtained with the double-THGEM, the correlation between these two quantities is much lower than in the single-THGEM setup.
The interpretation of this result is discussed in Section \ref{sec:discussion}.

\section{Electron transfer efficiency across the G-THGEM}
\label{sec:etransfer}

\subsection{Experimental measurement}
\label{sec:ete}

\begin{figure}
    \centering
    \includegraphics[width=0.7\columnwidth]{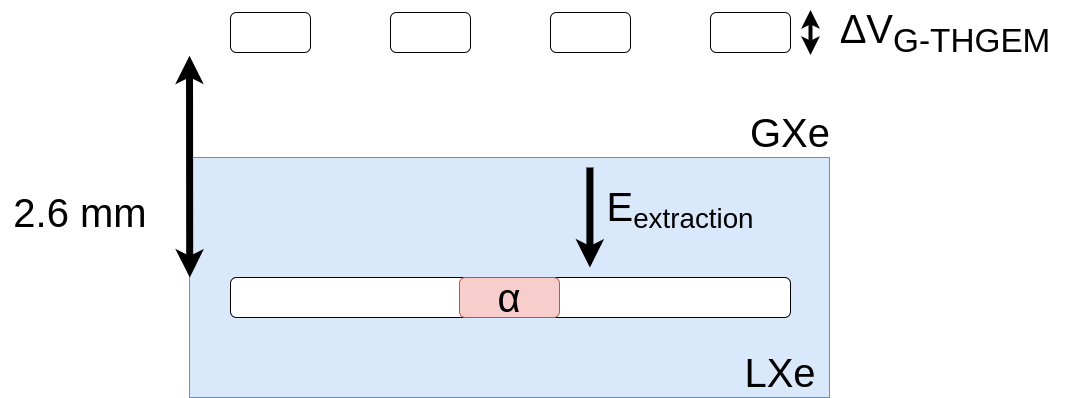}
    \caption{Schematic drawing of the setup used for the G-THGEM electron transfer efficiency measurement.}
    \label{fig:ete_setup}
\end{figure}

Due to the high electric field necessary to extract electrons into the gas phase efficiently \cite{Xu:2019dqb}, the focusing efficiency of electrons into the G-THGEM holes is compromised.
In order to estimate this effect, a dedicated measurement was performed.
The setup, depicted in Figure \ref{fig:ete_setup}, consisted of an \am source held in a stainless steel frame, immersed in LXe, and a bare THGEM electrode positioned in the gas phase with the liquid-gas interface located in between the two.
Ionization electrons released in the liquid are extracted to the gas phase under \eextraction and focused into the THGEM holes.
The measurement is performed in stationary mode, recording the currents on the THGEM top and on the source, with the electron transfer efficiency (ETE) defined as the ratio between the two:

\begin{equation}
    ETE = \frac{I_{top}}{I_{source}}.
    \label{eq:ete}
\end{equation}

\begin{figure}
    \centering
    \includegraphics[width=0.7\columnwidth]{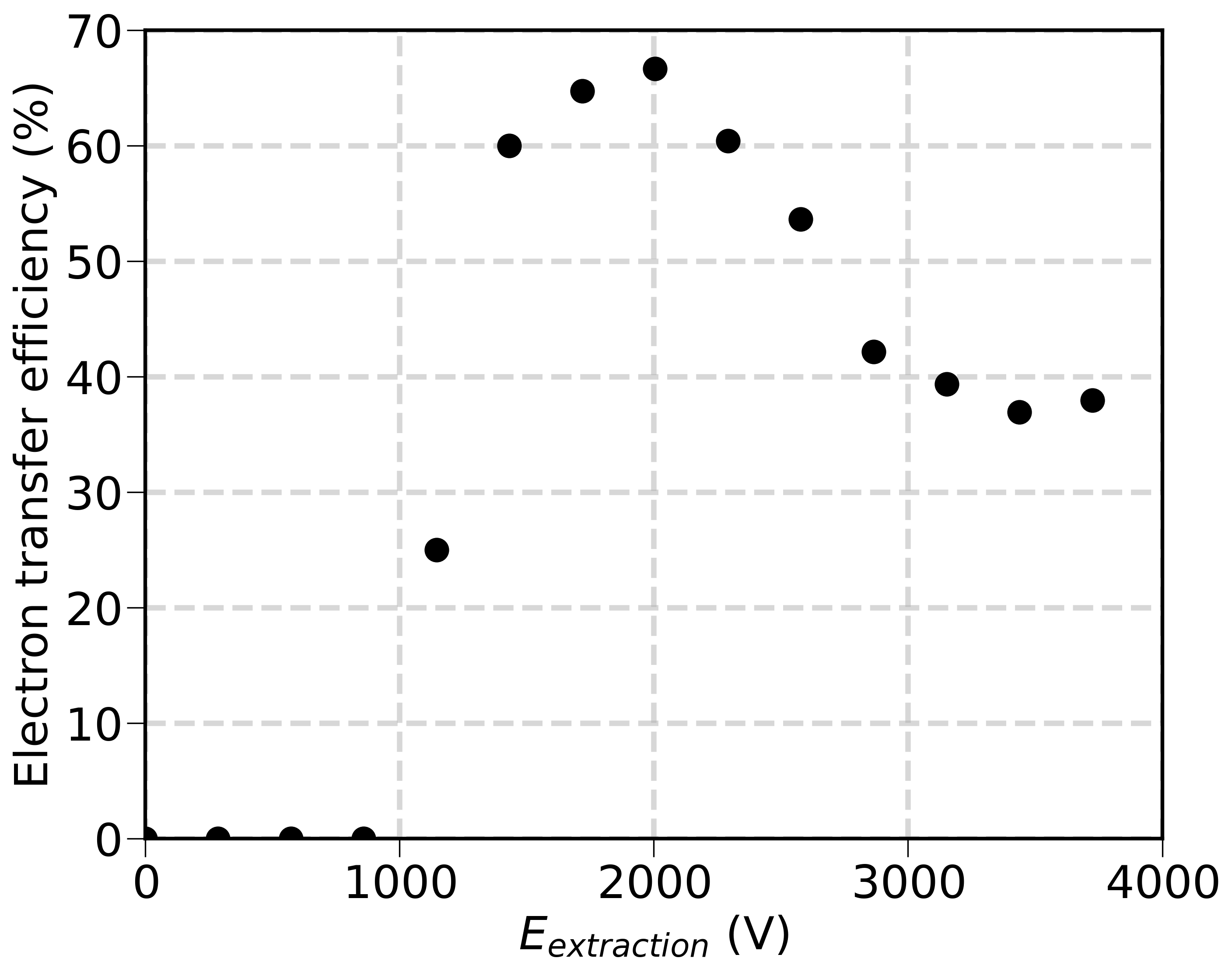}
    \caption{Estimated electron transfer efficiency across the G-THGEM in Figure \ref{fig:ete_setup} as a function of the extraction field for \dvgthgem = 1.6 kV. The maximum efficiency (67 \%) is measured at \eextraction = 2 kV/cm.
    }
    \label{fig:ete_gthgem}
\end{figure}

Figure \ref{fig:ete_gthgem} shows the ETE as function of \eextraction for a THGEM voltage of \dvgthgem~=~1.6~kV.
We observe that the optimal value ($\approx$67\%) occurs at \eextraction~=~2~kV/cm.
For a discussion on possible explanations and improvements on this result, the reader is referred to Section \ref{sec:discussion}.

\subsection{THGEM alignment calculations}
\label{sec:alignment}

The electron transfer efficiency results presented in Section \ref{sec:ete} must be interpreted as the average of the electron transfer efficiency over the active area of the \am alpha source.
However, in a cascaded LHM electrons reaching the G-THGEM do not have a homogeneous distribution in the transverse plane after traversing the L-THGEM. 
This raises the question of whether the alignment between the two THGEM electrodes plays an important role in the detector's performance.
To explore this effect, we performed finite-element calculations in COMSOL Multiphysics\registered.
We implemented a geometry equivalent to that of the right panel of Figure \ref{fig:setups} assuming two extreme cases: perfect alignment between the holes of the two THGEM electrodes and their total misalignment (plane rotation of 60$^\circ$).
From the simulation, we obtain the field lines, which are taken as a proxy for the electron trajectories.

\begin{figure*}
    \centering
    \includegraphics[width=0.487\textwidth]{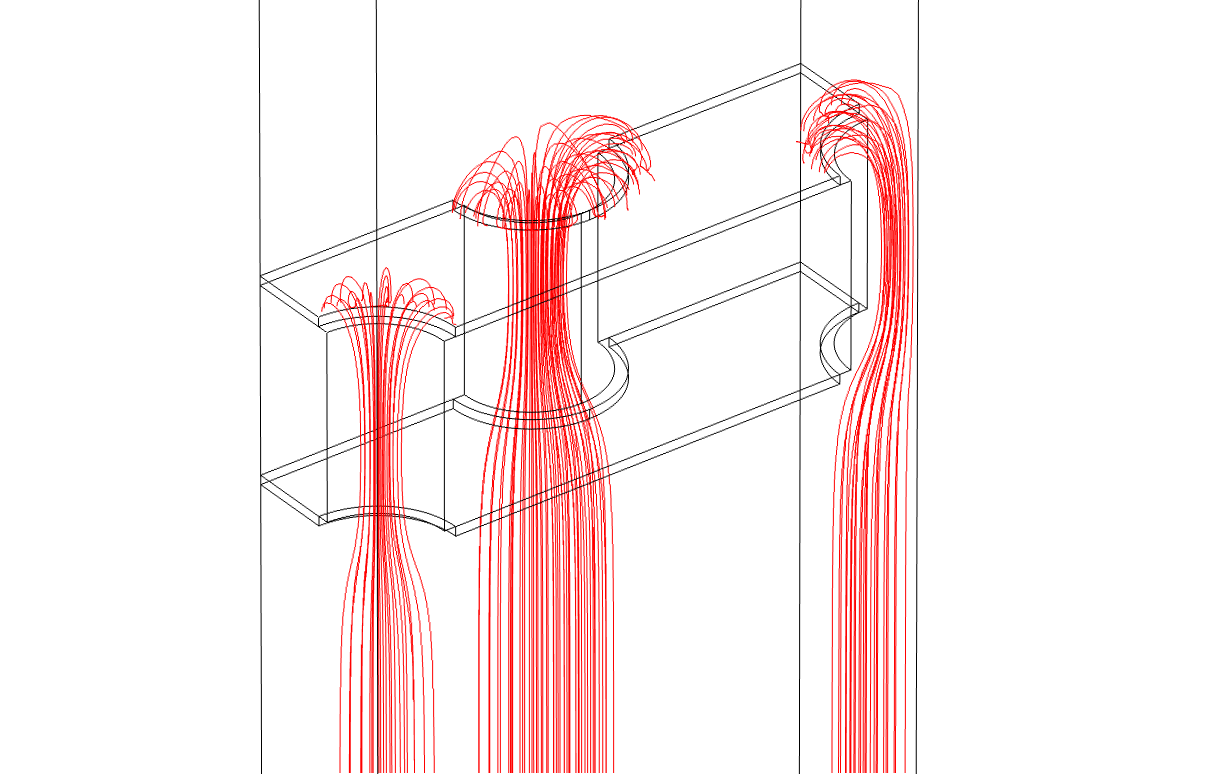}
    \includegraphics[width=0.487\textwidth]{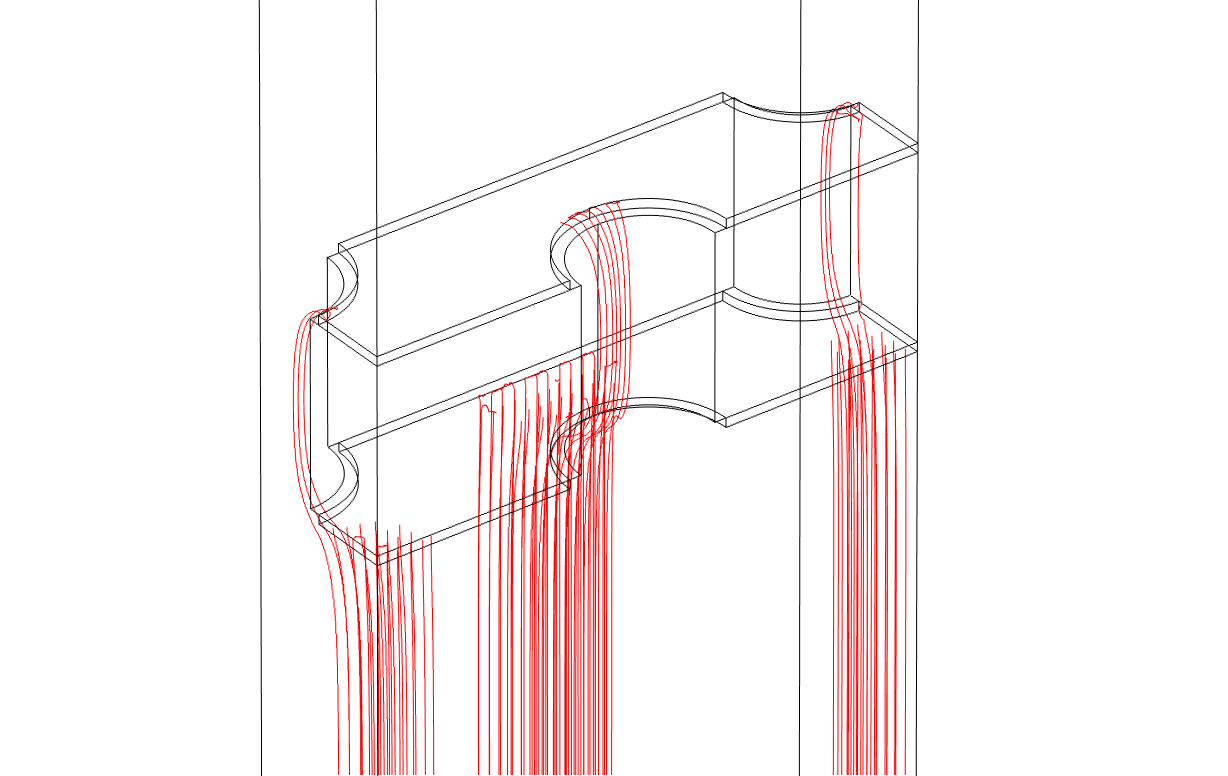}
    \includegraphics[width=0.49\textwidth]{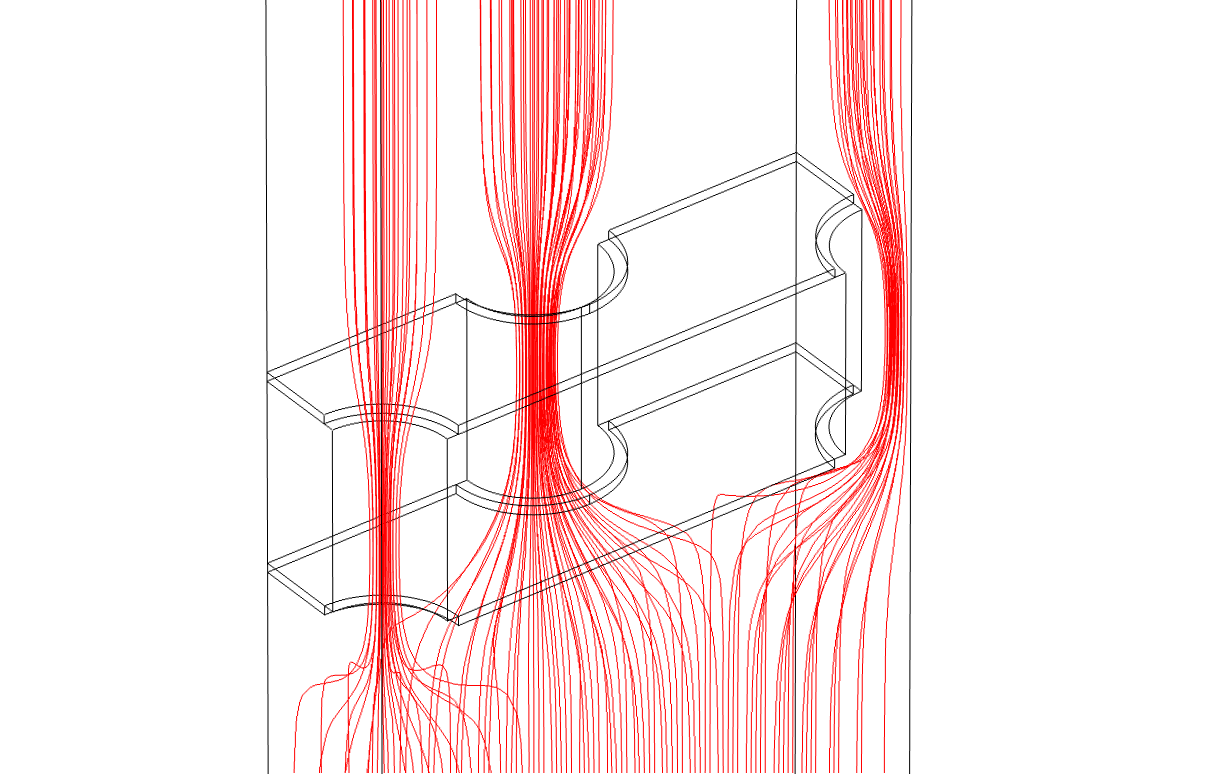}
    \includegraphics[width=0.49\textwidth]{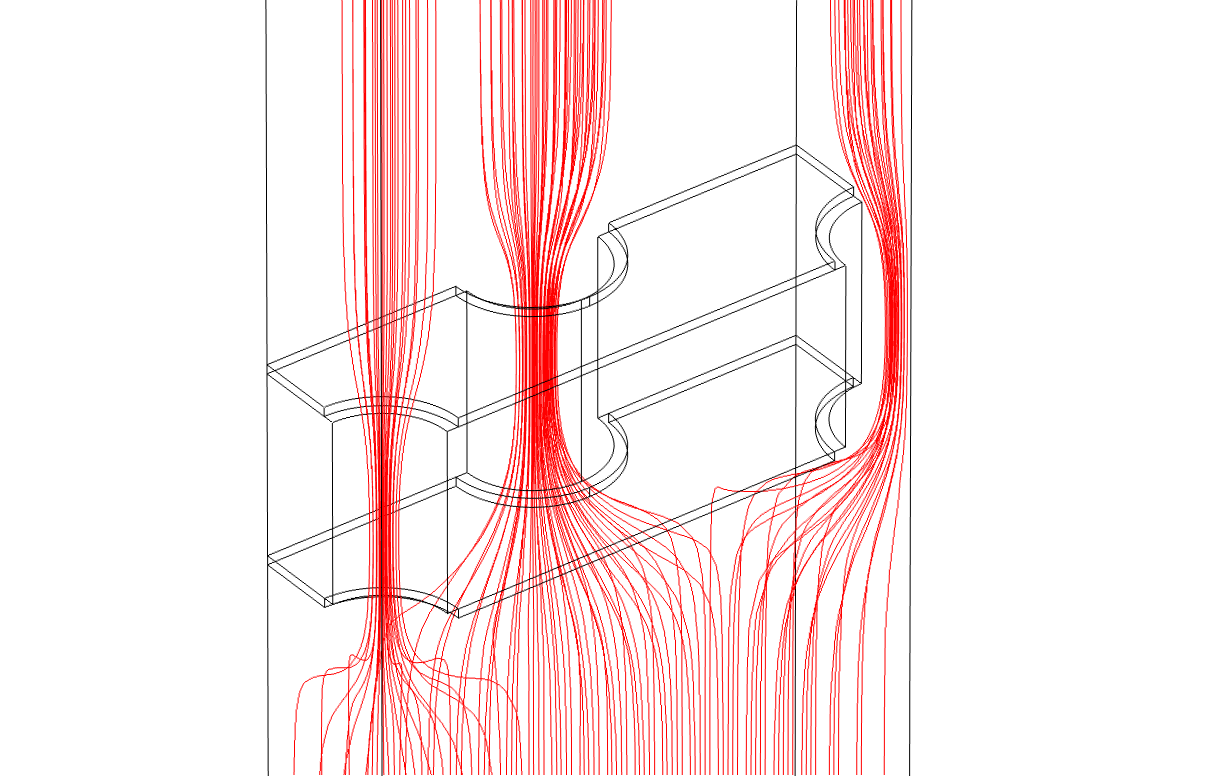}
    \caption{3D field lines calculated in COMSOL for perfectly aligned (left) and perfectly misaligned (right) THGEMs using electrodes with the same parameters as in our experiments. The bottom and top rows display a zoomed-in view around the L- and G-THGEMs respectively (the actual gap between the L- and G-THGEMs in much larger than shown here). While the field lines reach the top surface of the G-THGEM when both electrodes are aligned, almost all field lines end on the bottom surface of the G-THGEM when they are completely misaligned.}
    \label{fig:alignment}
\end{figure*}

Figure \ref{fig:alignment} shows the 3D field lines in a unit cell for the case of perfect electrodes' alignment (left) and total misalignment (right) configurations.
The bottom and top rows are zoomed-in views around the L- and G-THGEM electrodes respectively.
In both cases, the field lines are perfectly focused into the L-THGEM holes.
Moreover, we observe that when the THGEM electrodes are perfectly aligned, all field lines reach the top face of the G-THGEM, which can be interpreted as a transfer efficiency close to unity.
However, when the THGEM electrodes are fully misaligned, the field lines end at the bottom surface of the G-THGEM resulting in a transfer efficiency close to zero.

In our setup, both electrodes were placed in random orientations as we did not anticipate this effect.
Therefore, we must assume a random misalignment between the two electrodes.
Under these conditions, the measured electron transfer efficiency across the G-THGEM (Figure \ref{fig:ete_gthgem}) could not be determined precisely.

\section{Discussion and conclusions}
\label{sec:discussion}

We report on the first operation of the cascaded LHM: a dual-phase implementation of a LHM-based detector concept.
It is a variant of the original bubble-assisted LHM concept in which radiation-induced electroluminescence occurred within a gas bubble trapped under a perforated electrode \cite{Erdal:2015kxa}.
The cascaded LHM was devised to solve the principal drawback of the bubble-assisted LHM, namely, a low PDE value.
This was attributed to a poor transfer efficiency across the liquid-gas interface.
The curvature and the confinement of the interface to the THGEM holes, combined with the different electron trajectories for S1' photoelectrons and S2 ionization electrons were proposed as the root causes for the unsatisfactory detector performance.
Here, we explored a different arrangement, in which the liquid-gas interface is unconfined and placed above the immersed THGEM electrode.

This concept could offer potential solutions to three of the main problems in upscaling the current generation of dark-matter detectors.
The deployment of a CsI photocathode within the liquid enables the unambiguous detection of S1 signals.
Indeed, photoelectrons originating from CsI are amplified, after extraction to the vapor phase, via EL generating a high number of photons per photoelectron.
While EL multiplication in vapor phase can occur in a parallel gap, the implementation of a G-THGEM electrode decouples the processes of electron extraction into gas and EL amplification.
This has two main benefits: (1) it allows the usage of more intense extraction fields, as sagging is not a concern, minimizing the effect of delayed electron emission; and (2) it confines EL generation to the vicinity of the G-THGEM holes, away from the liquid-gas interface.
This reduces its sensitivity to imperfections in the liquid-gas interface such as vertical displacements or eventual tilts.
Note that covering large areas with THGEM electrodes requires tiling them into modules.
This may introduces dead areas, hampering the detector performance.
However, this can be avoided by diverting field lines away from the THGEM-electrode borders, using appropriate strip electrodes.

\subsubsection*{S1'/S2 ratio and PDE}
We explored two different configurations: one with a mesh in the gas phase, similar to a standard dual-phase detector, and one with a THGEM electrode in the gas phase, that decouples the processes of electron extraction into gas and EL amplification.
In both cases, we observe a significantly higher S1'/S2 ratio than in the bubble-assisted THGEM LHM \cite{Erdal:2017yiu}, which also translates into a higher PDE, as discussed above.
The values obtained were rather similar in the two configurations, although they were systematically higher in the double-THGEM one.
This (higher S1'/S2 for the double-THGEM), however, can be attributed to the larger solid angle subtended by the CsI photocathode.
The higher S1'/S2 ratios observed in both current configurations suggests that electron losses in a liquid-gas interface confined to the THGEM electrode holes is indeed the main factor responsible for the poorer PDE obtained in the bubble-assisted LHM.

From the S1'/S2 ratio measurements, we derived maximum PDE values of 3.7~\% and 1.5~\% for the single- and double-THGEM configurations respectively.
The PDE measured in the double-THGEM configuration is significantly lower than in the single-THGEM one.
We attribute this difference to a poor electron transfer efficiency across the G-THGEM electrode, as discussed below.
For a electron transfer efficiency across the G-THGEM close to unity, the PDE of the of the single- and double-THGEM configurations are expected to be equal.

Despite the improvement with respect to the bubble-assisted LHM, the currently obtained PDE values are significantly below expectation.
Eq. \ref{eq:pde} allows to estimate the PDE for a given configuration.
The effective quantum efficiency, \qeeff, is the average QE across the photocathode area.
This value can be obtained from the relation between the CsI QE in LXe and the electric field \cite{APRILE1994328} and the calculation of the local electric field strength on the photocathode's surface using COMSOL.
We estimate \qeeff~=~18.4\% and \qeeff~=~16.4\% at \edrift~=~238~V/cm and \edrift~=~920~V/cm, respectively.
Combining it with \epsone~=~0.9 \cite{Martinez-Lema:2023qox}, \epsext~=~0.9 \cite{Xu:2019dqb} and (in the double-THGEM configuration) \epsgas~=~0.4 (high \eextraction values in Figure \ref{fig:ete_gthgem}), we expect PDE~=~14.9\% and PDE~=~5.3\% for the single- and double-THGEM configurations, respectively.
These values are $\sim$3.75 times higher than those measured in this work.
Of the factors entering Eq. \ref{eq:pde}, \epsone was measured recently under very similar conditions \cite{Martinez-Lema:2023qox} and \epsext cannot explain such discrepancy is spite of the variations observed in the literature \cite{Xu:2019dqb}.
In addition, the fact that we observe roughly the same discrepancy factor in the single- and double-THGEM configurations indicates that the uncertainty in \epsgas is not to sufficient to justify this disparity.
Thus, \qeeff stands as the likely source of the deviation between the experimental measurement and the predicted value.
As detailed above, this number is obtained from finite-element calculations in COMSOL and a experimental measurement of the QE of CsI photocathodes in LXe.
We consider these calculations to be accurate, which leaves the QE value in each experiment as the only factor to account for this disagreement.
It is worth noting, however, that the QE value reported in \cite{APRILE1994328} has been reproduced independently by us \cite{lidine_2019, lidine_2021} with photocathodes of similar QE in vacuum, thereby challenging the attribution of the observed discrepancy solely to the QE measurement.
Further research is necessary to clarify this deviation.

\subsubsection*{Comparison between single- and double-THGEM configurations}
We observed that the double-THGEM configuration provides higher EL light yield compared to the single-THGEM one.
The maximum observed S2 signal of $3.8 \cdot 10^4$ PEs corresponds to 5.1 PEs per drifting electron.
This is roughly a factor 2 lower than that obtained by XENON1T \cite{XENON:2017lvq}.
We attribute it to a compromised electron extraction to the gas phase combined with their inefficient electron focusing into the G-THGEM holes as discussed below.

The double-THGEM configuration features also a better energy resolution, as shown in Figure \ref{fig:eres_comparison}.
However, the improvement in energy resolution can be partially explained by the increase in photoelectron statistics.
The energy resolution trend up to \dvgthgem~=~600~V can be attributed to the increased light yield.
However, there is no further improvement above this value, indicating that other effects are coming into play.
One possibility is that the focusing of electrons into the THGEM holes does not improve significantly above this value and that the main contribution to the energy resolution are the fluctuations in the number of electrons traversing the THGEM holes.
This can be attributed to the THGEM misalignment discussed in Section \ref{sec:alignment}.
Based on the results of our COMSOL calculations, for a large misalignment between the two THGEM electrodes, the number of field lines focused in the THGEM holes does not increase significantly with \dvgthgem, which supports this hypothesis.

Nonetheless, we consider the ability to tune the light yield independently from the extraction field an advantage with respect to the single-THGEM configuration.

The observed S2 signals in the double-THGEM configuration have a faster rise time and are shorter than those obtained with the single-THGEM one.
This can potentially be useful for improved time-separation between neighboring vertices along the drift axis.

We observed a stronger correlation between the S1' and S2 signals in the single-THGEM than in the double-THGEM setup.
Since S1' is a product of the primary scintillation and S2 from the ionization electrons, these two quantities should, in principle, be anti-correlated.
Both S1's and S2s result from electroluminiscence; therefore, any correlation between them can be assumed to originate from this process.
Thus, we attribute the positive correlation to variations in the liquid-gas interface.
Since the interface is not perfectly still --- e.g. being affected by vibrations, ripples or bubble formation with a characteristic time of tens of ms --- different events may occur under a slightly different transfer gap widths.
The single-THGEM configuration is more sensitive to these gap variations as the EL light is produced uniformly between the interface and the mesh.
On the other hand, the double-THGEM configuration is not as sensitive, since light production occurs predominantly in the vicinity and within the G-THGEM holes.

\subsubsection*{Electron transfer efficiency across the G-THGEM}
The electron transfer efficiency across the G-THGEM electrode was crucial for the understanding of our results.
We performed a direct measurement using a dedicated setup, representing the average electron transfer efficiency across the G-THGEM electrode.
The result pointed to significant losses in the extraction-to-gas and amplification stages.
We found that for the G-THGEM electrode used in this experiment, the optimal transfer efficiency value ($\sim$68\%) occurred at a very \eextraction ($\sim$2~kV/cm).
For the field configurations used in other measurements, a transfer efficiency of $\sim$40\% was measured.
Moreover, finite element calculations in COMSOL allowed us to identify the alignment between the two THGEMs as another critical factor in the optimization of the transfer efficiency.
These calculations indicate that the transfer efficiency varies drastically from close to zero (total misalignment) to near unity (perfect alignment).
As our assemblies did not include any element to ensure electrode alignment, we could not estimate the precise transfer efficiency and assumed the value obtained in the dedicated measurement.

Besides electrode alignment, there are other potential ways for improving the detector characteristics.
One possibility is to deploy perforated electrodes (THGEM, GEM and others) of different parameters in the liquid and vapor phases. These (e.g. hole diameter and spacing) should be chosen to maximize the effective area of the CsI photocathode (in the immersed electrode), as well as the electron and photoelectron collection and transfer efficiency through the holes - under conditions allowing for maximal electron extraction from the liquid.
The L-THGEM electrode could be replaced by other GEM-based electrodes, as demonstrated in \cite{Erdal:2017yiu} where the single-conical GEM emerged as the optimal electrode, offering a PDE 3-fold higher than that of the THGEM.
On the other hand, COMSOL calculations show that a G-THGEM electrode with larger holes provides better transfer efficiency and therefore, might be more suitable for this purpose.
However, using different THGEM geometries can incur into further alignment problems and a poor transfer efficiency.
Another improvement could come from adding electron-focusing strips between holes, similar to the the efficient ion-blocking concept described in \cite{LYASHENKO2009116}.

The results presented in this work encourage further research of these detector concepts, focused primarily on improving energy resolution and photon detection efficiency.



\appendix

\section*{Acknowledgements}
The authors would like to thank Dr. Sergei Shchemelinin from the Unit of Nuclear Engineering of the Ben-Gurion University of the Negev for his assistance with the CsI evaporation.

\bibliographystyle{elsarticle-num} 
\bibliography{src/bibliography}

\end{document}